\def\gs{\mathrel{\raise1.16pt\hbox{$>$}\kern-7.0pt
\lower3.06pt\hbox{{$\scriptstyle \sim$}}}}
\def\ls{\mathrel{\raise1.16pt\hbox{$<$}\kern-7.0pt
\lower3.06pt\hbox{{$\scriptstyle \sim$}}}}

\documentstyle[editedvolume,numreferences]{crckapb10}
\newcommand{\stt}{\small\tt}

\begin{opening}
\title{Radio Spectra and NVSS Maps of Decametric Sources}
\subtitle{}

\author{O. VERKHODANOV}
\institute{Special Astrophysical Observatory, Nizhnij Arkhyz, Russia 357147}
\author{H. ANDERNACH}
\institute{Dpto.\,de Astronom\'\i a, IFUG, Apdo.Postal 144, Guanajuato, Mexico}
\author{N. VERKHODANOVA}
\institute{Special Astrophysical Observatory, Nizhnij Arkhyz, Russia 357147}
\author{N. LOISEAU}
\institute{INSA; ESA IUE Observatory, Apdo.\,50727,\,E--28080\,Madrid,\,Spain}

\end{opening}

\runningtitle{DECAMETRIC RADIO SOURCES}

\begin{document}
\ifx\href\undefined\else\errmessage{Don't use hypertex!}\fi

\vspace*{-9.0cm}
\begin{footnotesize}\baselineskip 10 pt\noindent
Proc. {\it Observational Cosmology with the new Radio Surveys}, Tenerife,
Spain, Jan.\,13--15, 1997 \\
eds.~~M.\,Bremer, N.\,Jackson \& I.\,P\'erez-Fournon, Kluwer Acad.\,Press,
in press
\end{footnotesize}
\vspace*{8.0cm}

\begin{abstract}
We constructed radio spectra for $\sim$1400 UTR-2 sources and find that
46\% of them have concave curvature.  Inspection of NVSS maps of 700
UTR sources suggests that half of all UTR sources are either blends of 
two or more sources or have an ultra-steep spectrum (USS).
The fraction of compact USS sources in UTR may be near 10\%.
Using NVSS and the Digitized Sky Survey(s) we expect to double the 
UTR optical identification rate from currently $\sim$19\%.
\end{abstract}

The UTR-2 catalogue (hereafter ``UTR'') of 1754 radio sources \cite{Brau94}
covers $\sim$30\% of the sky 
($\delta_{50}$=[--13$^{\circ}$...+20$^{\circ}$; +41$^{\circ}$...+60$^{\circ}$])
at $\nu$=10--25 MHz and angular resolution 
$\theta_{\rm RA}\times\theta_{\delta} = 40'\times40'$sec($\delta$--49.7$^{\circ}$)
at 16.7 MHz.
We intend to improve the optical identification (ID) fraction 
(of now $\sim$19\%) for this lowest-frequency source 
sample presently available, in order to look for
new or rare species of sources.

To construct radio spectra of UTR sources we used ``CATS''\cite{And97}
to extract all 6C,\,MIYUN,\,B3,\,MSL,\,TXS,\,MRC,\,WB92,\,PMN,\,87GB 
and GB6 sources within 40$'$ from the UTR location.
The ``raw'' spectra given by these fluxes were refined using 
computer charts of source locations around UTR positions.
All counterparts from TXS, GB6 and PMN within circles of 1$'$ radius were 
considered one source.  Groups of sources lying further apart 
were assigned separate spectra, each with the UTR flux as upper limit.
We fitted spectra of 1525 radio counterparts to 1401 UTR sources
with either straight (S), convex (C$^{-}$), 
or concave (C$^{+}$) curves in the lg\,$\nu$--lg\,S plot.
We found 47/46/7\% 
of type S/C$^{+}$/C$^{-}$, while other authors found partitions of
46/5/39\%, ~23/6/42\%, and 22/6/25\% respectively, at decametric \cite{RBC73}, 
intermediate \cite{Herb92}, and high frequencies \cite{Kuhr81}.
The high fraction of C$^{+}$ spectra in UTR sources may be partly intrinsic 
due to source selection at very low $\nu$, and partly be caused 
by blends of steep- and flat-spectrum sources in one UTR beam. 

We inspected 1.5\,GHz maps of the NVSS \cite{Cond97}
for $\sim$700 UTR sources with {\stt RA}$<8^{\rm h}$.
The maps were 72$'$--134$'$ on a side, depending on UTR positional errors
and HPBW at 16.7\,MHz.
16\% of the maps were not useful due to significant holes in coverage.
No dominant source was seen in the innermost half of 66\% of the useful
maps, suggesting that (a) the decametric flux may be a blend of unrelated 
sources within the UTR beam or that (b) the UTR source may be faint 
at 1.5\,GHz due to a steep spectrum.  Apart from many complex Galactic plane
sources the NVSS maps revealed a few examples of previously unknown, 
radio- and optically-faint, and very extended (10$'\ls$LAS$\ls$18$'$) 
FR\,I type radio galaxies, some at low galactic latitude.

We had useful NVSS maps for 43 UTR sources without a radio counterpart 
listed in \cite{Brau94}. 40 of these lack a dominant NVSS source, i.e.
the UTR signal may be a blend of unrelated sources. 
One of the other 3 is a USS cD galaxy (A193, $\alpha$=2.8, S$\sim\nu^{-\alpha}$),
thus we plan to look for USS sources among the 40 former ones as well.

There are 90 UTR sources identified with 92 PKS sources in \cite{Brau94}, 
42 of which without optical ID.  Today only 3 more optical IDs are known 
from \cite{PKSCAT} or NED. 
Of the 19 NVSS maps we had for the 42 unidentified UTR-PKS sources,
only 3 showed a dominant 1.5\,GHz source, again suggesting
confusion in the UTR catalogue.


We looked for UTR objects in two USS samples with VLA maps.  
One set \cite{Rhee96} (4C sources with $|\delta_{50}|<4^{\circ}$, 
$\alpha_{365}^{178}>$0.9 and LAS$_{\rm TXS}<$30$''$)
should be near or above the UTR limit, but
out of 146 UTR--4C associations with $|\delta_{50}|<4^{\circ}$
listed in \cite{Brau94}, only 29 coincide with objects from \cite{Rhee96}. 
This may be due to a combination of 
(a) source sizes $>$30$''$; 
(b) sources near the 4C limit being missed by UTR; 
(c) spectral flattening below $\sim$100\,MHz (as observed in compact 
steep-spectrum sources) or 
(d) TXS having lost some flux from these extended 
sources, implying a spectrum flatter than assumed by \cite{Rhee96}.
The other published USS sample \cite{Rott94} has $\sim$350 sources
in the UTR region. Among the 44 safe coincidences with UTR radio counterparts
we find a lower fraction of point sources and more doubles (7\% and 75\%) 
compared to 24\% resp.\ 58\% in the full sample of 350 USS sources.

26\% of the NVSS maps of the UTR sources without optical ID 
in \cite{Brau94} show one (or occasionally two) sources dominating the field.
If these are the true IDs and optically bright enough, we may 
hope to raise the ID rate of UTR sources to $\sim$40\%
using only NVSS, FIRST and the Digitized Sky Survey(s). \\[-2.ex]

We thank S.\,Trushkin for advice on Galactic sources, N.\,Zabavskaya 
for typing part of UTR, and the {\stt SkyView} team at GSFC for the NVSS map server.
O.V.\,thanks the Russian Foundation of Basic Research for support (under
grant 96-07-89075 for CATS). ~H.A.~acknowledges a travel grant from the 
conference organizers.\\[-4.ex]

\end{document}